\documentstyle[aps,prb,preprint]{revtex}

\tighten
 
\begin{document}
\draft
\preprint{over12.tex - 25.8.98}
 
\title{Correlation and symmetry effects in transport
 through an artificial molecule}
 
\author{F. Ram\'{\i}rez\cite{FR}$^a$, E. Cota$^b$ and S. E. Ulloa$^c$}

\address{$^a$Posgrado en F\'isica de Materiales, Centro de
Investigaci\'on Cient\'ifica y de Educaci\'on Superior de Ensenada,
B.C., M\'exico \\ $^b$Centro de Ciencias de la Materia Condensada --
UNAM,  Ensenada, B.C., M\'exico \\ $^c$Department of Physics and
Astronomy and Condensed Matter and Surface Sciences Program, Ohio
University, Athens, OH 45701-2979} 

\maketitle 

\begin{abstract}
 Spectral weights and current-voltage characteristics of an artificial
diatomic molecule are calculated, considering cases where the dots
connected in series are in general different. The spectral weights
allow us to understand the effects of correlations, their connection
with selection rules for transport, and the role of excited states in
the experimental conductance spectra of these coupled double dot
systems (DDS). An extended Hubbard Hamiltonian with varying interdot
tunneling strength is used as a model, incorporating quantum
confinement in the DDS, interdot tunneling as well as intra- and
interdot Coulomb interactions.  We find that interdot tunneling values
determine to a great extent the resulting eigenstates and corresponding
spectral weights.  Details of the state correlations strongly suppress
most of the possible conduction channels, giving rise to effective
selection rules for conductance through the molecule.  Most states are
found to make insignificant contributions to the total current for
finite biases.  We find also that the symmetry of the structure is
reflected in the I-V characteristics, and is in qualitative agreement
with experiment.
 \end{abstract}

\narrowtext
 
\section{Introduction}

A semiconductor quantum dot or `artificial atom' is an electronic
device defined on the nanometer scale.\cite{kastner,other-revs} This
system usually arises when a homogeneous two dimensional electron gas
($2DEG$) generated at the interface between layers of semiconductor
structures is laterally confined by electrostatic, mechanical, or
other means. An artificial atom is characterized by a strong
quantization of the electronic motion in all three spatial
dimensions. This means that the spectrum of the electrons is discrete
with separation between levels given by a characteristic value,
$\Delta$.  On the other hand, the extremely low capacitance (both self
and mutual) achieved in these nanostructure systems, due to the small
sizes and compact geometries of the arrangements, produce a relatively
large charging energy of $U=e^2/C \simeq 1$ meV.\@ In most
semiconductor structures in this regime, one finds typically that $U >
\Delta$, frequently differing by an order of magnitude or more.

In a typical `lateral' transport structure, \cite{scott} which
consists of a quantum dot coupled via tunnel barriers to two
reservoirs (source and drain) and a back gate, the number of electrons
can be controlled at will, starting from a small number of electrons
(or none) in the dot.  The charging of the $N$-electron atom with an
additional single electron can be done by changing the back gate
voltage, as it controls the depth of the local potential-well holding
the electrons.  The charging takes place when the chemical potential
of the emitter electrode equals the `local' chemical potential of the
atom, by providing enough energy for the system to receive a particle.
Since a large energy $U$ is required, this gives rise to the {\em
Coulomb blockade} ($CB$) of transport whenever this energetic
condition is not met.  The $CB$ is in fact the mechanism for the
surprisingly strict control of the charge in the quantum dots, and
their denomination as artificial atoms with a well-defined number of
electrons at a given set of gate voltages.  As a consequence of $CB$,
only one electron at a time can tunnel into the quantum dot for high
or wide tunneling barriers, and one observes oscillations of the
differential conductance as a function of the back gate voltage, which
controls the equilibrium charge of the dot, every time the electron
population in the dot changes.  \cite{averin,beenakker,meir} This
in-plane geometry has been explored extensively both experimentally
and theoretically.

In a different geometry, an ingenious device that uses a {\em
capacitor} where electrons tunnel between a metallic layer and
discrete quantum levels of the confined structure has been studied
recently by several groups. \cite{other-revs,ashoori} This sensitive
device monitors small capacitance peaks as a function of voltage
across the structure every time an electron tunnels into the dot.
This method of single electron capacitance spectroscopy has allowed
researchers to monitor the intricate behavior of the many-particle
states produced as function of external magnetic
fields. \cite{ashoori} Other interesting techniques used to
investigate properties of artificial atoms include far-infrared
spectroscopy, \cite{meurer} which explores excitations of these
artificial atoms, and recent `vertical' transport experiments in novel
gated multi-quantum well structures. \cite{Tarucha,Guy} Together,
these experimental probes provide fascinating insights into the
properties of these artificial atoms and molecules, in a similar way
to what atomic and molecular physics yield, although the different
energies and variable electron number are unlike anything possible in
those systems.

The conductance in transport spectroscopy through a well-defined
artificial atom is strongly affected by the Coulomb blockade, as
described above, \cite{genref} and is a clear manifestation of charge
quantization.  Notice that when the voltage difference between the
source-drain leads is small, one is in the linear-response regime.
Here, one can see the collection of differential conductance peaks in
terms of the so-called {\em addition spectrum}, i.e., the series of
energy values required to add one electron to the system.  This is
given by the chemical potential in the leads (and equal to one another
in this linear regime), $\mu_N \equiv E_{N,1} - E_{N-1,1}$, where
$E_{N,1}$ is the ground state energy of the $N$-electron artificial
atom.  As the back gate (or some other neighboring gate) voltage is
shifted, it produces successive conductance peaks in the transport
experiment. \cite{Beenakker-orig} Therefore, one can say that the
linear regime provides a direct measure of the ground states of the
system.

On the other hand, in the {\em nonlinear} transport regime, at finite
drain-source bias voltages, additional conductance peaks are observed
which reflect the presence and nature of the excited states of the
artificial atoms for a given particle number.
\cite{weis,johnson,foxman} McEuen {\em et al.} \cite {mceuen} have
realized transport spectroscopy on single dots and carefully analyzed
the role of excited states versus source-drain bias and magnetic
fields.  In the case of single quantum dots, exhaustive studies of the
excited states pointed out that a large number of the available states
do not contribute to the conductance, signaling the existence of
selection rules {\em for transport}.  In fact, a number of theoretical
works demonstrated that indeed unusual selection rules are required to
account for the observed suppression of the fine structure.  These
selection rules appear due to strong correlations in the electron
eigenstates and corresponding eigenfunctions.
\cite{palacios,weinmann,pfann} The appearance of strict spin selection
rules and/or those related to the orbital motion have been associated
with the many-particle nature of these states and provide a natural
explanation of the experimental data.
 
Our goal in this article is to understand how discrete energy levels,
electron-electron interactions and symmetry affect the spectrum in an
artificial diatomic molecule (coupled quantum dots), and how this is
reflected in the linear and nonlinear transport characteristics.  This
will be especially important for the strongly-correlated few-electron
regime, as it is widely expected that increasing carrier number or
concentration ends up making the quantum dot not too dissimilar from a
classical polarizable droplet (at zero magnetic field).  Arrays of
quantum dots have been modeled to study the addition spectra and
conductance. \cite{stafford,klimeck,charlie} Notice also that
transport measurements have been reported for arrays of two or more
dots connected in different geometries.  These artificial molecules
have conductance peaks which split as a function of interdot
interaction, \cite{vandervart,waugh} and show interesting charging
diagrams, be it in a series, \cite{waugh,blick} or parallel
connection.  \cite{wharam,parallel-guy} Linear and non-linear
transport experiments conducted on two coupled dots in series indicate
that as interdot tunneling is turned-on, this interaction allows
charge to distribute throughout the system and controls the evolution
from a two-dot system to a larger dot. \cite{blick,livermore,crouch}
Beautiful direct evidence of a fully-developed coherent-resonant
`molecular state' (in terms of the classical `symmetric/antisymmetric'
or `bonding/antibonding' quantum mechanical states) has recently been
presented by Blick {\em et al}. \cite{blick,newblickprl} These authors
have focused on the study of a coherent molecular state than can be
found in the charging diagram of the double dot system.  This charging
diagram is constructed by varying `top' and `back' gate voltages in
the linear transport regime, and `triple points' were identified where
the device could be used as an electron pump.

The evolution of the differential conductance as a function of
interdot tunneling for the series-connection has been treated
theoretically for a symmetric double dot system by Kotlyar {\em et
al}., \cite{rozita} combining a step-well model for the confinement
potential of the system used in Ref.\ \onlinecite{crouch}.  They used
a Mott-Hubbard model to describe the electronic interactions, and
obtained excellent qualitative agreement with the measured currents in
the nonlinear regime.

Here we study an artificial diatomic molecule that is a simple coupled
array of two quantum dots connected in series.  We consider the general
case where the two dots are not identical (both the `symmetric' and
`asymmetric' cases), similar to the system in Ref.\ \onlinecite{blick}.
 We model the system with an extended Hubbard Hamiltonian which takes
fully into account the interaction between quantum dots in a real
system: interdot tunneling interaction defined in a typical lateral
structure by tunable gates, and the intra- and interdot Coulomb
repulsion.  We apply the analysis of the spectral weights (`overlaps')
following Ref.\ \onlinecite{pfann}, for the few-electron eigenstates of
the quantum system.  The Hamiltonian allows us to calculate exactly the
entire energy spectrum of this multi-particle system by numerical
diagonalization, as well as the full eigenfunctions of the system.  The
current through the molecule is determined to a great extent by the
spectral weights of the states involved in the transitions in the dots,
which also describe the electronic correlations in the system. 
Regarding only sequential tunneling, the total current through the
artificial molecule incident from the left reservoir can be written
explicitly as (with a similar expression for transport through the
right barrier) \cite{kinaret}
 \begin{equation}
 I = -e \sum_{\alpha \alpha'} \tilde \Gamma^L_{\alpha
 \alpha'}[P((N-1),\alpha') f^L _{\alpha \alpha'}-P(N,\alpha)(1-f^L
_{\alpha \alpha'})] \,. 
 \label{i-eq}
 \end{equation} 
 The Fermi distribution function $f^L _{\alpha \alpha'}=f_{FD}(\Delta
E_{\alpha \alpha'}-\mu_L)$, characterizes the occupation of the
electron levels in the left reservoir (with chemical potential $\mu
_L$). Here the resonant energy $\Delta E_{\alpha \alpha'}=
E_{N,\alpha}-E_{N-1,\alpha'} $, is the difference between the energy
of an $N$-particle state $\alpha$,  $|N, \alpha \rangle$, and an
$(N-1)$-particle state, $\alpha'$,  $|N-1, \alpha' \rangle$.  The
probability $P(N,\alpha)$ of finding the quantum molecule in the
$N$-particle state $\alpha$ will deviate from its equilibrium value for
a given drain-source voltage. Its dependence on the tunneling rate 
$\tilde \Gamma^L_{\alpha\alpha'}$ is well described by kinetic
equations. \cite{Beenakker-orig,kinaret,DP-FKP}
 The corresponding tunneling rate  $\tilde \Gamma^L_{\alpha \alpha'}$
depends on the single-electron tunneling rate  $\Gamma^L_n$ for an
electron traversing the system in the state $n$, and the details of the
multi-particle states.  Since the energy (or $n$) dependence of
$\Gamma_n^L$ is weak and/or monotonic, we further conclude that the
tunneling rate is dominated by the intrinsic spectral weight, so that
$\tilde \Gamma_{\alpha \alpha'}^L = \gamma^L S^L_{\alpha \alpha'}$,
where $\gamma^L$ is a smoothly energy dependent single-particle
tunneling rate, and the overlap or spectral weight is
 \begin{equation}
 S^L_{\alpha \alpha'} = \sum_{n \in L} |\langle N, \alpha|C_n^{\dagger}
 |N-1, \alpha' \rangle|^2 \, .
 \label{s-eq}
 \end{equation} 
 This quantity describes the correlations in the system, and its
contribution to the current (I--V) characteristics proves to be
dominant in determining the salient features measurable in experiments.
The spectral weights govern the tunneling probability because they
describe the overlap between the $N$-electron state $\alpha$, and the
compound state built by an incoming electron with quantum number(s) $n$
added to the $(N-1)$-electron state $\alpha'$.  For a system of
uncorrelated electrons this overlap will be either one or zero between
any two states, by definition, in an orthogonal basis.  However,
electron correlations result in overlaps much less than unity, as the
correlations built into the states severely limit the possible
`conduction channels', and the tunneling probability is consequently 
reduced considerably. 

Our study here of the overlap matrix elements not only gives us
insights into the physical process behind the selection rules, but
also allows us to explore the general properties of the current
characteristics to be measured in these systems.  The aim of this work
is to investigate the effect of interdot tunneling interaction and
interdot Coulomb repulsion on the spectral weights and current-voltage
characteristics through a double dot system (DDS).  Given recent
experiments with dots with markedly different sizes, we also study the
effect of this structural asymmetry on the state correlations and
ensuing transport properties.  This asymmetry, typically implemented
with top gate arrangements, provides an additional parameter which
allows exploration of the correlations in the system.

In the Hubbard approach we use here, we find that the interdot
tunneling interaction has a direct effect on the spectral weights and
I--V characteristics, since it controls the possible delocalization of
the wavefunction and effectively regulates the correlation of the
different states.  The spectral weights critically depend on the
number of electrons $N$ because interactions change every time an
electron enters the system, and the number of channels increases
rapidly with $N$.  We find also that the structural asymmetry is most
evident in the I--V characteristics for small interdot tunneling, but
present even for relatively well-connected dots in the DDS.

\section {Model}

We use the extended Hubbard Hamiltonian,
 \begin{equation}
 \hat H = \sum_{j\alpha} \epsilon_{j\alpha} \hat C_{j\alpha}^\dagger 
\hat C_{j\alpha} - \sum_{\alpha \beta i j} (t_{\alpha \beta}  \hat
C_{i \alpha}^\dagger \hat C_{j \beta} + h.c.) \\ + {1\over2} \sum_j U_j
\hat n_j (\hat n_j - 1) + \sum_{ij} V_{ij}\hat n_i  \hat n_j \, ,
 \label{h-eq}
 \end{equation}
 where the parameters take into account the different types of
interactions. Here $C_{j\alpha}^\dagger$ and $\hat C_{j\alpha}$ are
creation and annihilation operators, $\hat n_j$ is the electron number
operator at site $j$, and $\epsilon_{j\alpha}$ are the confined energy
levels of the $\alpha^{th}$ state in the $j^{th}$ quantum dot; these
levels are assumed to be equally spaced with separation $\Delta_j$ (as
appropriate for a local harmonic oscillator confinement potential
which should be a good description of typical `lateral' dots). $U_j$
is the on-site Coulomb repulsion for the $j^{th}$ quantum dot,
$V_{ij}$ is the interdot repulsion, and $t_{\alpha \beta}$ is the
tunneling matrix element between the single particle states $\alpha$
and $\beta$ in the respective neighboring dots.  Kotlyar and coworkers
have presented a parameterization of the classical capacitance matrix
elements in terms of the Hubbard Hamiltonian quantities. \cite{rozita}
Some of the details will change from their square-well potential to
our harmonic oscillators, and these depend on gate geometries and
other structural features.  In either case, one would obtain intrinsic
Hubbard parameters with characteristic values of $U_j \simeq 1$ meV,
and $U_j \gg V_{ij} \simeq 0.1$ meV, in the typical GaAs structures
used in experiments.

The parameter $t_{\alpha \beta}$ is perhaps the most sensitive to the
specific gate implementation and applied gate voltages.  In fact, the
interdot barrier transparency has been used superbly to control the
overall interdot conductance in the experiments of Crouch {\em et
al}., \cite{crouch} and Blick {\em et al}., \cite{blick} to name a few
groups. \cite{tjerk} These tunneling parameters effectively control
the correlations between states in the DDS, by limiting the
wavefunction overlaps.  It is the well-known competition of this
tunneling with the Coulomb interactions that determine the details of
correlations in the states. \cite{stafford}

The specific values of the tunneling matrix elements depend on how the
interdot barrier is formed and modeled, so that $t_{\alpha \beta}$ can
be assumed to be given by a gaussian distribution (in energy
difference) that simulates the expected decreasing coupling between
levels that are not resonant or nearly so. \cite{santafe}  In order to
evaluate the effect of the interdot coupling differences, we compare
two different regimes. On the one hand, the case of a diagonal matrix,
$t_{\alpha \beta}=t \delta_{\alpha \beta}$, describes tunneling between
aligned states only (likely the case for high/wide barriers).  On the
other hand, the case of a constant distribution given by  $t_{\alpha
\beta}=t$, where tunneling between all states is allowed, give us two
opposite coupling regimes.  This latter case can be used to describe
the strong tunneling regime resulting when the interdot barrier is low
and/or narrow.  For a dot of diameter $d=100$ nm in a GaAs/AlGaAs
heterostructure, the charging energy $U \simeq 1$ meV, which greatly
exceeds the thermal energy $k_BT$ at the characteristic dilution
refrigerator temperatures of $ \simeq 0.1$ K, so that it is safe to
assume that these devices  work in the quantum regime, $k_BT < t <
\Delta_j < U$. In this description, we may use spin orbitals and the
spin overlap contribution can be considered, \cite{weinmann} especially
for finite magnetic fields, but we choose to model the artificial
molecule as a system of spinless fermions for simplicity. This
restriction can be clearly relaxed, but given the typically much
smaller Zeeman splitting, we do not expect that our conclusions would
be drastically changed at these temperatures and for typical
structures.

The procedure we follow is to solve the extended Hubbard Hamiltonian
(\ref{h-eq}) in the particle number representation by direct
diagonalization to obtain the eigenvalues and eigenvectors for the
system with $N$ electrons, and use Eq.\ (\ref{s-eq}) to calculate the
spectral weights. The system wavefunctions are expressed in the local
orbital representation, and we find then how the creation operator
$C_n^{\dagger}$ transforms the state $|N-1, \alpha' \rangle$, for
example.  Any of these states is a linear combination of local
orbitals with coefficients (probability amplitudes) which describe the
state fully.  As the electron enters the DDS, it delocalizes into a
complex molecular electronic state, $C_n^{\dagger} |N-1,\alpha'
\rangle$.  The projection of this new state of the molecule over the
state $|N, \alpha \rangle$ gives us information about that
delocalization which is a product of the interplay between the hopping
reducing confinement and the Coulomb interaction, which effectively
suppresses tunneling. The analogy with chemistry, describing our
system as a covalently bonded artificial molecule, gives us a deeper
insight into the processes taking place here. \cite{blick} Notice,
however, that the repeated and sequential particle addition to the
molecule via transport through the leads is clearly unlike any process
in atomic or molecular physics, as mentioned in the introduction.

 We find that in the  strong tunneling  and interacting regime (highly
correlated system) most of the  spectral weights take values near zero
and only some specific channels dominate  the spectra as occurs for
single dots. \cite{pfann}  The consequences of these strong  and
effective selection rules for the current through the system are
calculated using Eq.\ (\ref{i-eq}), or the equivalent symmetrized
expression (taking into account the tunneling from the left and the
right explicitly)
 \begin{equation}
 I = \frac {e}{2} \sum_{\alpha \alpha'} P(N-1,\alpha') \left[ 
 \tilde \Gamma _{\alpha \alpha'}^R f^R_{\alpha \alpha'} 
 - \tilde \Gamma_{\alpha \alpha'}^L f^L_{\alpha \alpha'} \right]
 + P(N, \alpha) \left(f^R_{\alpha \alpha'} - f^L_{\alpha \alpha'}
 \right) \, ,
 \label{i-full}
 \end{equation}
 which we use in all calculations below.  In this equation, the
factors $P(N,\alpha)$ are the probabilities of having the system with
$N$ electrons in the state $\alpha$.  These can be obtained from the
solution of rate equations, as discussed in the literature,
\cite{Beenakker-orig} for an accurate evaluation of all the limiting
rates during the conduction process.  Here, we assume for simplicity,
that these probabilities are well described by a superposition of two
equilibrium distribution functions determined by the chemical
potentials at each reservoir, so that $P = (P_R + P_L)/2$, where
$P_{L/R} = \exp (-\beta (E_{N,\alpha} - N \mu_{L/R})) / Z(\mu_{L/R})$.
Here, $Z$ is the Gibbs distribution function for each of the
reservoirs at chemical potential $\mu_L$ and $\mu_R$. Although this
independent `feeding' of the DDS by each reservoir is only an
approximation, it turns out that it is not too far from the full
solution of the rate equations, except for large biases, and whenever
the overlaps change drastically with energy. \cite{DP-FKP} In this
expression, we have also added the appropriate bias and gate voltage
dependence to the energy spectrum, so that $E_{N,\alpha} =
E_{N,\alpha}^0 - eN(c_G V_G + c_B V_{DS})$, where $E_{N,\alpha}^0$ are
the eigenvalues of the Hamiltonian (\ref{h-eq}), and the constants
$c_G$ and $c_B$ are proportional to the capacitance between the dots
and the gates defining the voltages.  As a typical example of
non-identical dots, we take a constant value of $\lambda = c_B / c_G =
2/3$, while one would expect $\lambda = 1/2$ in a symmetric structure.

As a finite bias $V_{DS}$ is applied to the DDS, one is in the
nonlinear transport regime, and the left and right reservoirs are
offset from each other by $eV_{DS}= \mu_L - \mu_R $.  When
sufficiently large bias voltage $V_{DS}$ is applied, new channels are
open for electron conduction and the overlaps measure the probability
for single electron tunneling through each channel.  Since the
spectral weights $S_{\alpha \alpha'}$ provide the information for the
current-voltage characteristics, and in the nonlinear regime the
energy scale of interest is $ \Delta E_{\alpha \alpha'} \propto
eV_{DS} $, we analyze these weights over varying energy intervals
$\Delta E_{\alpha \alpha'}$ to identify the channels
$(\alpha,\alpha')$ that contribute to the current in that
interval. Thus, we have the possibility of identifying the particular
channel that contributes to the current at a given voltage and proceed
to compare with the experimental results.  In this regime, to
calculate the current with Eq.\ (\ref{i-full}), we must take into
account all channels in the appropriate energy window, but our
calculation shows that only a rather small number of them contribute
significantly to the transport rate via Eq.\ (\ref{s-eq}).

\section{Results and Discussion} 

In what follows, we will measure all energy parameters in the
Hamiltonian in terms of the local repulsion $U \simeq 1$ meV,
characteristic of typical systems.  To provide contrast for the
different regimes, we consider here two cases: (a) the case of
symmetric quantum dots, where the harmonic-oscillator level spacing
and intradot Coulomb interaction are the same in each site, i.e.,
$U_1=U_2=1$ and $\Delta_1=\Delta_2=0.3$; (b) the `asymmetric' case,
where the two quantum dots in the molecule are not the same, and the
structural parameters are different.  As an example, we take
$\Delta_1=0.3$, $\Delta_2=0.2$ and $U_1=1$, $U_2=0.8$, corresponding
to a larger dot with the index 2.  Notice that since the dot 2 is
assumed larger (with size $L_2$), both the harmonic confinement
($\propto L_2^{-2}$), and the local repulsion term ($\propto
L_2^{-1}$) yield smaller values than for the small dot in site 1.  We
should also mention that inclusion of a finite and reasonable interdot
interaction $V_{12}$ ($\lesssim U/10$) yields rather small energy
shifts in the energy level spectrum (similar to a slight rescaling of
the value of $U$), and negligible effects in the spectral weights, in
general.  In what follows, and without loss of generality, we present
results with $V_{12}=0$.

\subsection{Spectral weights}

In Fig.\ \ref{1}, we show results for the spectral weights $S_{\alpha
\alpha'}$ for the {\em symmetric} double dot system as a function of
the energy difference $\Delta E_{\alpha \alpha'}$ between the states
involved in the transition, corresponding to the case where the number
of electrons goes from $N=2$ to $N=3$ for different intensities of
tunneling coupling.  In all the figures, we identify with filled
circles the channels $(\alpha,1')$, that represent transitions between
{\em all} possible states $|N,\alpha \rangle$ of $N$ electrons and the
{\em ground} state $|N-1,1' \rangle$ of $(N-1)$ electrons; empty
circles denote all other pairs. Figure \ref{1}(a) illustrates that
without tunneling ($t=0$) the electrons in the system are totally
uncorrelated and, correspondingly, the overlaps are either zero or
one.  Figures \ref{1}(b),(c) reveal the connection between electronic
correlations and tunneling measured by $S_{\alpha \alpha'}$ in the
weak tunneling regime, for different coupling models.  In \ref{1}(b)
we use the diagonal tunneling matrix $t_{\alpha \beta}= t
\delta_{\alpha, \beta}$ with $t=0.1$. As $t$ is gradually increased
one obtains progressively smaller values of $S_{\alpha \alpha'}$ for
most state pairs, and only a chosen few are non-zero (notice large
number of circles on the horizontal axis).  This general behavior is
also obtained for different $t_{\alpha \beta}$-matrix coupling, even
if the details of the suppressed transition pairs change somewhat. In
Fig.\ \ref{1}(c), we couple all single particle states between dots
with an energy-dependent gaussian distribution with a maximum at
$t=0.1$ on resonance.  Notice the rather similar behavior to
\ref{1}(b).  The extreme regime of strong coupling is explored by
taking a constant distribution $t_{\alpha \beta}=t$. To compare
results we use again $t=0.1$, and see in Fig.\ \ref{1}(d) that a
signature of this most correlated system is a strong suppression of
the weights $S_{\alpha \alpha'}$, and a clear prevalence of the
transitions involving the ground-states (highest spectral weights
occur for the lowest energy filled circles in \ref{1}(d)), and
low-lying excitations of the DDS. \@ As most of the channels have low
spectral weights, their contribution to the current will be small in a
transport experiment, according with Eq.\ (\ref{s-eq}). Only the few
channels with large spectral weights would give origin to discernible
peaks in the differential conductance traces, as we will see in the
next section.  In particular, in Fig.\ \ref{1}(d), the largest
contribution comes from the transition involving the ground states
from $N=2$ to $N=3$, suggesting that nonlinear conductance features
would be quite small at finite bias.  This observation is verified
later when we actually calculate I-V diagrams and is in qualitative
agreement with experiments.  \cite{crouch}

Figure \ref{2} presents typical results for overlaps in the symmetric
DDS case for the sequential addition of electrons, from $N=1$ to
$N=5$.  The number of particles in the system obviously modifies the
interactions and, as a consequence, the eigenfunctions generate
different spectral weights for each channel with the addition of
electrons. We use here $t=0.1$ for all pairs in this system, a
strongly correlated case.  We observe that the correlations in the
artificial molecule are different for the same interval $\Delta E
\propto V_{DS}$ for different $N$, since interactions readjust every
time an electron enters the system.  The symmetry of the artificial
molecule leaves as signature on the spectral weights that the channel
$(1,1')$ ends up having always the maximum overlap.  The results are
different in the asymmetric case.  As will be shown below, there are
many channels $(\alpha, \alpha')$, that are directly related to the
excitation spectra of the artificial molecule, which provide major
contributions via large values of the overlaps.

If we calculate the spectral weights for an asymmetric molecule, we
observe the effects of the dot asymmetry for different values of
tunneling amplitude.  For example, we show in Figs.\ \ref{3}(a) and
(b) spectral weights $S_{\alpha \alpha'}^L$ for an electron that
enters the system from the left in the transition $N=2 \rightarrow
3$. Figures \ref{3}(c) and (d) show the corresponding results for an
electron incident from the right, $S_{\alpha \alpha'}^R$. For the same
channels in the same interval $\Delta E$, we obtain in general that
$S_{\alpha \alpha'}^R \neq S_{\alpha \alpha'}^L$, as one would expect
that the single-particle level asymmetry would carry over to the
many-particle states.  The overall reduction of the overlaps is
evident when $t$ increases, as one electron is added to the system
from either left or right.  At the same time, we notice a big
difference with the symmetric molecule. In the asymmetric case many
channels corresponding to transitions between excited states have
large overlaps even in the strong tunneling regime, and {\em even
larger} than the (1,1') ground state transitions for $S^L_{\alpha
\alpha'}$ in Fig.\ \ref{3}(b).

Increasing the number of electrons sequentially in the asymmetric case
gives the values of the overlaps shown in Fig.\ \ref{4} for incidence
from the left (emitter side for positive $V_{DS}$). Here, only the
contribution from transitions ($\alpha,1'$) are shown (from
$E_{N-1,1}^0$ to $E_{N,\alpha}^0$). In the insets, the corresponding
overlaps for incidence from the right are shown. In each case, the
largest contribution comes from the transition between ground states,
and is larger for incidence from the right (left) for transitions from
even (odd) to odd (even) $N$, as expected from the asymmetry
considered (larger dot on the right).  These transitions involve
states which are weighted predominantly on either of the dots,
increasing the overlap that comes from a given side, and decreasing
the other.

The previous analysis illustrates two characteristics of the overlaps.
First, they provide selection rules in transport spectroscopy that
allow us to explain why the flow of current from the left can be lower
or larger for a particular channel as discussed above.  The important
point is that for any channel in question, we can analyze the
calculation and offer an explanation of their contribution to the
current. The second point is closely related with the physical process
of tunneling. This coupling mixes single electron states and builds up
a molecular state where electrons are correlated.  Thus, many electron
wave-functions of the artificial molecule contain information about
`bonds' between quantum dots. Tunneling provides a kind of bonding
interaction and, as described, this affects the spectral weights in a
non-trivial way.  In the local-orbital approximation, we can
understand that upon operating with the creation operator
$C_n^\dagger$, we create an $N$-particle state,
$C_n^{\dagger}|N-1,\alpha' \rangle$, which is in general not one of
the eigenstates of the system, $|N,\alpha \rangle$, but is rather
represented as a linear combination of basis vectors in the
$N$-particle Hilbert space.  If there is a dominating $(N-1)$-particle
state in a given spectral weight, say the $j^{th}$ state, the overlap
will be given by $S_{\alpha,\alpha'}(n) \simeq |\lambda_j|^2|\mu_k|^2
\, $.  Here, $|\lambda_j|^2$ is the probability that the system with
$(N-1)$ electrons is in the $j^{th}$ basis state.  In a similar way,
$|\mu_k|^2$ represents the probability that the artificial molecule
occupies the $k^{th}$ basis state with $N$ electrons.  If we assume
that both states correspond to maximum (or minimum) probability, this
qualitative simplification gives us another approach to explain high
or low spectral weights in terms of probabilities associated with
basis states.  For example, we can explain the values of overlaps for
channel $(1,1')$ in Fig.\ \ref{1}(b) this way. We calculate the
spectral weight and obtain $|\lambda_{1'}|^2$=0.9997 for an electron
incoming from left or right. The operation $C_1^{\dagger}|N-1, 1'
\rangle$ gives a vector that is projected over the ground state $|N,
1\rangle$. This state has a maximum occupation probability
$|\mu_1|^2$=0.4984, so that the overlap is given by $|\lambda_{1'}|^2
|\mu_1|^2 \simeq 0.5$. The exact value obtained from Eq.\ (\ref{s-eq})
is 0.4996.  Notice that this simple analysis gets complicated rather
quickly as $t$ increases, since many more occupation probabilities
$|\lambda_j|^2$ need to be considered to build up the given state
$|N-1,\alpha' \rangle$, i.e. electrons delocalize with increasing
tunneling interaction. Spectral weights reflect this delocalization
and measure the corresponding correlations in the system. If the
interactions in the system are such that the system is strongly
correlated, more states participate in the overlaps, but their
contributions come with different phases.  This gives rise to a strong
suppression of the available transport channels, so that only a few
contribute significantly to the current.

\subsection{Current-voltage characteristics}

We can use the values of the overlaps obtained in the previous section
to calculate the current through the system as a function of gate
voltage $V_G$ and source-drain voltage bias $eV_{DS}=\mu_L-\mu_R$, via
Eq.\ (\ref{i-full}). $V_{DS}$ is specified by the chemical potentials
of emitter and collector electrodes, while sweeping the gate voltage
$V_G$ through positive values shifts down the electrostatic potential
of the $N$-electron system in the DDS.  For the symmetric case, we
present in Fig.\ \ref{5} results for the current for a range of values
of interdot tunneling, in a grey-scale contour plot, where dark
corresponds to small current (lighter shades indicate higher $|I|$
values, with sign equal to that of $V_{DS}$). The temperature for
these calculations was set at $k_B T=0.01 \, (\simeq 120$ mK), an
order of magnitude smaller than the mean level spacing.  The general
characteristics of these plots have been analyzed for single
\cite{weis,stewart} and double \cite {crouch} quantum dot systems.
Notice that here we plot current, and not the differential conductance
typically plotted, as a function of both $V_G$ and $V_{DS}$, providing
similar physical information (sample differential conductance traces
are discussed below).

In these plots, the current in the linear regime, at $V_{DS} \simeq 0$
shows $CB$ steps corresponding to changes in the ground state of the
coupled dots from the $N$ to the $N+1$ electron configuration (the
`addition spectrum').  The differential conductance traces would show
$CB$ peaks in the linear regime, separated by the actual charging
energy required to add one electron to the system.  As $V_{DS}$
increases, excited states of both configurations become accessible
near each $CB$ peak, providing new tunneling channels through the
double dot. This results in broadening of the $CB$ peaks to form
multiple peak structures enclosing `Coulomb diamonds' (appearing
darkest in Fig.\ \ref{5}), corresponding to $CB$ regions of zero
conductance and fixed electron number, as indicated in each diamond.
The lines defining the diamond edges correspond to transmission
`resonances' or alignment of the ground states of the DDS with source
or drain Fermi levels.  Lines parallel to the edges and away from the
$CB$ diamonds correspond to transitions involving excited states of
the quantum molecule. For positive $V_{DS}$, we identify resonances
parallel to the negative (or positive) slope Coulomb diamond edges as
unoccupied $DDS$ levels in resonance with the source (or drain) Fermi
level, i.e., $\mu_L$ (or $\mu_R$).

For the ideal symmetric double dot system for weak tunneling in Fig.\
\ref{5}(a), we obtain Coulomb blockade regions corresponding to
`even-even' double-dot ground states in which each dot has the same
number of electrons, and increasing the gate voltage adds electrons
{\em in pairs} to the system, as the symmetric and antisymmetric
quantum mechanical states are nearly degenerate for small $t$, and the
local interaction dominates.  The charging energy and corresponding
$CB$ diamonds are quite large (in comparison with those in other
panels, see below).  As $t$ increases, Fig.\ \ref{5}(b), and the
energy splitting between ground states becomes significant, we see
that the current steps split, producing two different size
diamonds. The smaller diamonds signal the $N$ odd states, as one would
expect that the split-pairs in a bonding/antibonding picture would be
far away in energy from the next.  This is intuitively expected for
single-particle states mixed by a weak tunneling matrix element.  That
it is also the case for a multi-particle state with several local
orbitals mixed-in by tunneling and Coulomb interaction is a somewhat
surprising result.  Moreover, Blick {\em et al}. have shown that this
mixing of many-particle ground states by parts is quite successful in
describing experimental data in a double-dot geometry. \cite{blick}

 As $t$ increases further, the two types of diamonds are nearly
identical, (Fig.\ \ref{5}(c), indicating that the states of the
individual dots are fully mixed into an overall single dot with
smaller charging energy (and then smaller $CB$ diamonds) without any
even-odd charging differentiation.  Basically identical results have
been nicely obtained in Ref.\ \onlinecite{rozita} by combining a
two-site generalized Hubbard model with a one-dimensional step-well
model for the confining double dot potential.  Their calculated
nonlinear transport characteristics are in excellent qualitative
agreement with experiment by Crouch et al., \cite {crouch} for which
they were designed.

In the case of asymmetric dots, recent experimental studies by Blick
{\em et al}. have presented the charging diagram of such series
structure in the linear regime, as discussed above. \cite {blick} Our
results for the asymmetric case (larger QD on the right) in the
nonlinear regime, are shown in Fig.\ \ref{6} again as a grey-scale
contour plot of the current in the $V_G$-$V_{DS}$ plane. Here we see a
structure of small and large diamonds, even at $t \simeq 0$, which
reflects the built-in asymmetry in the dots producing differences in
charging energy from even to odd $N$, as opposed to the ideal
symmetric case where we have full level degeneracy and charging by two
electrons at a time (see Fig.\ \ref{5}(a)). Also, note that I--V
traces are {\em substantially} different depending on the {\em sign}
of the polarization $V_{DS}$, since occupancy of states in the DDS in
the collector side differs from that on the emitter side.
Additionally, note that as in the symmetric case, small diamonds
increase in size and width as interdot tunneling increases, since
tunneling allows the extra electron to be shared between the two dots,
and tends to effectively reduce the structural asymmetry.  By the time
$t=0.2$ in this asymmetric molecule, one can hardly distinguish the
structural asymmetry any more, as shown in Fig.\ \ref{6}(c).  The
interdot tunneling has basically transformed the DDS into a single
larger dot, with nearly identical $CB$ diamonds for all $N$'s {\em
and} even similar excited-state structures.

From this discussion, it is clear that the most asymmetric situation
occurs whenever the quantum dots making the molecule are not identical
and the interdot tunneling is not too large.  To further illustrate
this, in Fig.\ \ref{7} we plot the current and the differential
conductance, $dI/dV_{DS}$, as a function of $V_{DS}$, for fixed values
of the gate voltage $V_G$ and interdot tunneling.  In the symmetric
case, Fig.\ \ref{7}(a), $V_G=1.09$ and $t=0.1$, while in the
asymmetric cases, Figs.\ \ref{7}(b) and (c), $V_G=0.87, t=0.1$ and
$V_G=1, t=0.01$, respectively.
 
 These values of $V_G$ are taken from Figs.\ 5 and 6, and correspond
to the charging energy for the transition $N: 2 \rightarrow 3$.
\cite{lastnote} The excited state symmetry is clearly observed in the
curves shown for both the current and its derivative, as can be seen
in \ref{7}(a).  The central peak in differential conductance is the
ground state to ground state contribution, and is clearly the most
important in all cases, although less so in the asymmetric structure.
The lateral peaks correspond to the contributions from excited states,
and we see that these start contributing at a smaller positive value
of $V_{DS}$ in the asymmetric case.  Notice that the large feature in
\ref{7}(b) at $V_{DS} \simeq -1$ corresponds to transitions involving
the ground states $N: 3 \rightarrow 4$ which have become accessible at
the finite bias.  However, at $\mid V_{DS} \mid < 1$, we see a number
of transitions via excited states, some of which have quite a large
value.
 
In the case of the effectively more asymmetric system, given its small
value of interdot coupling $t=0.01$, the differential conductance is
remarkably asymmetric, as shown in Fig.\ \ref{7}(c).  The large
feature at $V_{DS} \simeq -0.6$ is associated with the ground state
transition $N: 3 \rightarrow 4$ (just as above), while all the other
smaller peaks are related to excited states: The feature at $V_{DS}
\simeq -0.4$ is produced by an excited state of $N=3$ in the DDS,
while that at $\simeq -0.8$ is via a $N=4$ excited configuration.  On
the other hand, the two differential conductance peaks for $V_{DS} >
0$ are excited states of the $N=2$ configuration which make quite a
large contribution to the current and conductance.  Notice that other
transitions involving excited states (clearly seen in \ref{7}(b), or
in the corresponding spectral weights) are suppressed here.  Once
again, this is consequence of the subtle wavefunction mixing that
takes place for non-zero $t$.  We should point out that although
larger differences are apparent in Fig.\ \ref{4} for the spectral
weights for left- or right-incidence in asymmetric dots, this is not
carried over as sharply in the current or conductance curves.
Inspection of the symmetrized expression for the current, Eq.\
(\ref{i-full}), suggests that the addition of the appropriate terms
with $\Gamma^L$ and $\Gamma^R$ tends to de-emphasize these
differences.

\section{Conclusions}

Using an extended Hubbard Hamiltonian, which takes into account intra-
and inter-dot Coulomb interactions and variable interdot hopping, we
calculate the overlap matrix elements $S_{\alpha \alpha'}^{L/R}$ and
corresponding I-V characteristics for artificial diatomic molecules,
in both symmetric and asymmetric geometries.  The calculations are
performed in the weak and strong tunneling regimes, as the number of
electrons in the system increases from $N=0$, as a function of the
gate voltage.  The effects of interdot Coulomb interaction are found
to be small and equivalent to a minor rescaling of the local intradot
interaction, in agreement with Stafford {\em et al}. \cite{stafford}

It is found in all cases that only a few of the many channels
involving excited states of the DDS contribute to the current. In the
symmetric case, the largest contribution corresponds to the channel
involving the ground states of $N$ and $N+1$ particles. Indeed, the
contour diagrams of nonlinear current as a function of gate voltage
$V_G$ and source-drain voltage $V_{DS}$ (Fig.\ \ref{5}) show the
formation of primary and secondary diamonds ($CB$ regions) evolving as
interdot tunneling increases, in excellent qualitative agreement with
experiments with nearly-identical coupled
dots. \cite{livermore,crouch} In the asymmetric case, with the largest
dot on the right, we have to consider the difference in the overlaps
$S_{\alpha \alpha'}^L$ and $S_{\alpha \alpha'}^R$ for incidence from
left or right, respectively.  In the {\em strong} tunneling regime, we
find that in contrast to the symmetric case, there are several
channels involving excited states that contribute to $S_{\alpha
\alpha'}^L$.  On the other hand, the main contribution to $S_{\alpha
\alpha'}^R$ comes from the ground states of $N$ and $N+1$
particles. From the behavior of the overlaps in this case we can say
that the system is less correlated and that there are strong competing
effects between tunneling coupling and asymmetry.  These effects enter
in the calculation of the current, and the final influence of excited
states in the asymmetric case can best be appreciated in the diagrams
for the current in the {\em weak} tunneling regime of Figs.\
\ref{6}(a) and (b) or \ref{7}(c).  The experimental work by Blick et
al. \cite {blick} on asymmetric double dot structures was concerned
with the $V_{DS}=0$ regime.  We believe that analysis of finite bias
data in these structures should give unique insights into the dot
molecular states, and encourage experimental groups to test asymmetric
structures.  Finally, a detailed theoretical analysis of the
differential conductance as a function of `back' and `top' voltages,
as defined in the split-gate experimental setup of Ref.\
\onlinecite{blick} is possible with our present approach, and will be
presented elsewhere.

\acknowledgements 

This work was supported in part by CONACYT project 0078P-E9506 and
DGAPA-UNAM IN100895.  SEU is supported in part by US DOE grant no.\
DE--FG02--91ER45334.

\begin{figure}
 % \centerline{\epsfxsize=9cm \epsfbox{fig1.ps}}
 \caption{Spectral weights as a function of the energy difference
$\Delta E_{\alpha \alpha'}$ between states involved in the transition
$N: 2 \rightarrow 3$, for different interdot tunneling models: (a)
$t=0$, (b) diagonal $t_{\alpha \alpha'}=t\delta_{\alpha\alpha'}$, (c)
gaussian, (d) constant $t_{\alpha\alpha'} =t$.  Filled circles
represent channels ($\alpha,\alpha'=1$) for all $\alpha$.  Symmetric
DDS case. $U_1=U_2=U$; $\Delta_1=\Delta_2=0.3U$.}
  \label{1}
\end{figure}

\begin{figure}
 \caption{Effect of number of electrons $N$ on spectral weights in the
constant regime ($t_{\alpha\beta}=0.1$)  Symmetric molecule case.}
 \label{2} \end{figure}

\begin{figure}
 \caption{Spectral weights in asymmetric DDS case, with $U_1=1;
U_2=0.8; \Delta_1=0.3;$ and $\Delta_2=0.2$.  Electrons incident
from left ((a) and (b)) or right ((c) and (d)), with $t$ values as
shown, for the  $N:2\rightarrow 3$ transition.}
 \label{3}
\end{figure}

\begin{figure}
 \caption{Spectral weights for left incidence for transitions
involving the $(N-1)$-particle {\em ground state}, for $t=0.1$ in the
constant regime and $V=0$.  Insets show results for incidence from
right.}  \label{4}
\end{figure}

\begin{figure}
 \caption{Current as a function of gate voltage $V_G$ and source-drain
voltage $V_{DS}$, for different values of interdot tunneling: (a)
$t=0.01$, (b) $t=0.1$ and (c) $t=0.2$.  Symmetric DDS case.}
 \label{5}
\end{figure}

\begin{figure}
 \caption{Current in $V_G$-$V_{DS}$ plane for asymmetric DDS.  (a)
$t=0.01$, (b) $t=0.1$, and (c) t=0.2.  Notice asymmetry is nearly
absent in (c) but still clearly seen in the finite bias current
steps.}  \label{6}
\end{figure}

\begin{figure}
 \caption{Current and its derivative $dI/dV_{DS}$ as a function of
source-drain voltage $V_{DS}$ for (a) symmetric $t=0.1$; (b)
asymmetric $t=0.1$, and (c) asymmetric $t=0.01$ DDS.  Gate voltages
fixed as explained in text.  Conductance peaks away from $V_{DS}=0$
are produced by a few of the excited states.}  \label{7}
\end{figure}


\begin{references}
\bibitem{lastnote} Notice that for the calculations in Fig.\ 7 we have
used $\lambda=1/2$ to study only the effects of the state symmetry and
not those due to the asymmetry in potential drop across the DDS.
\end{references}
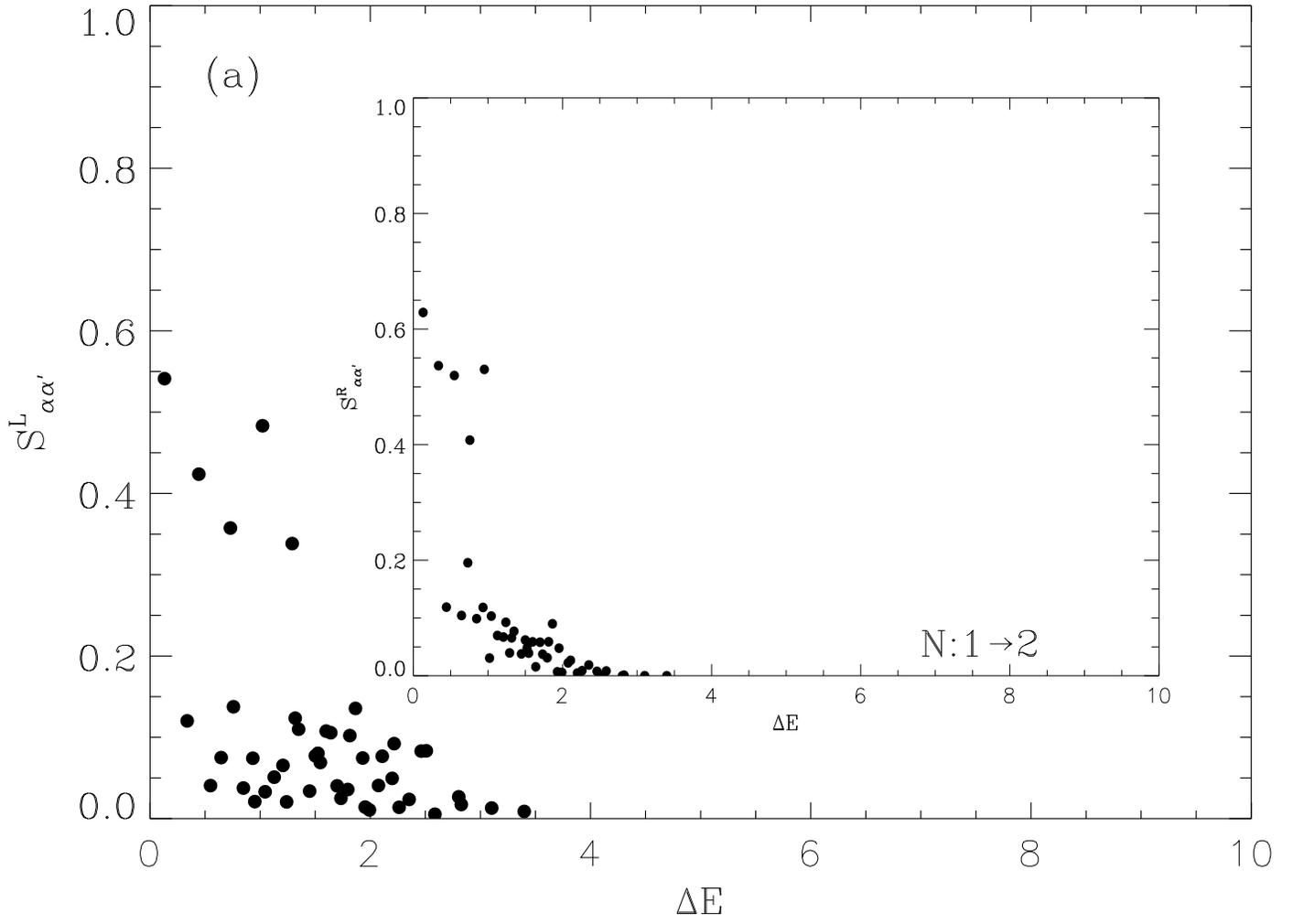
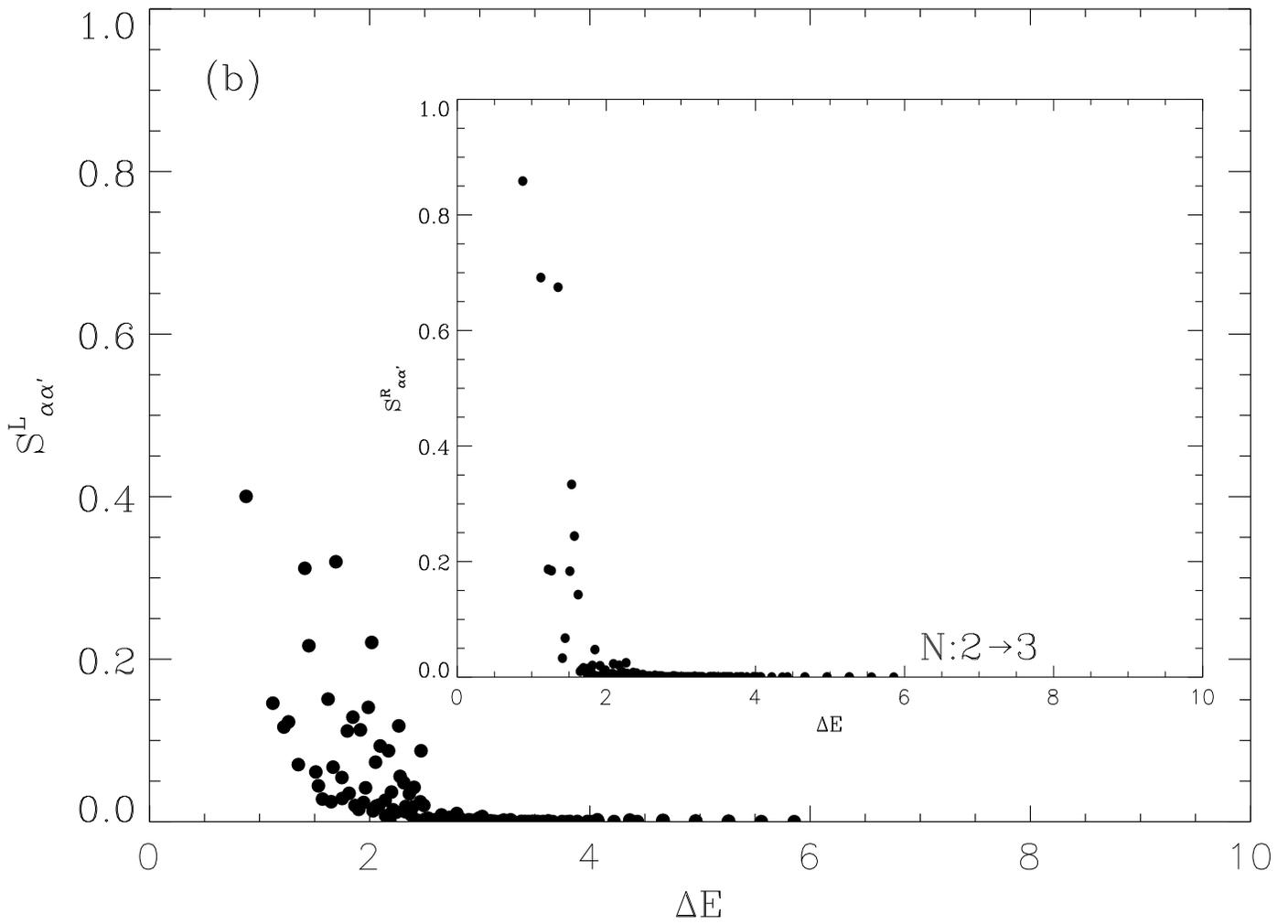
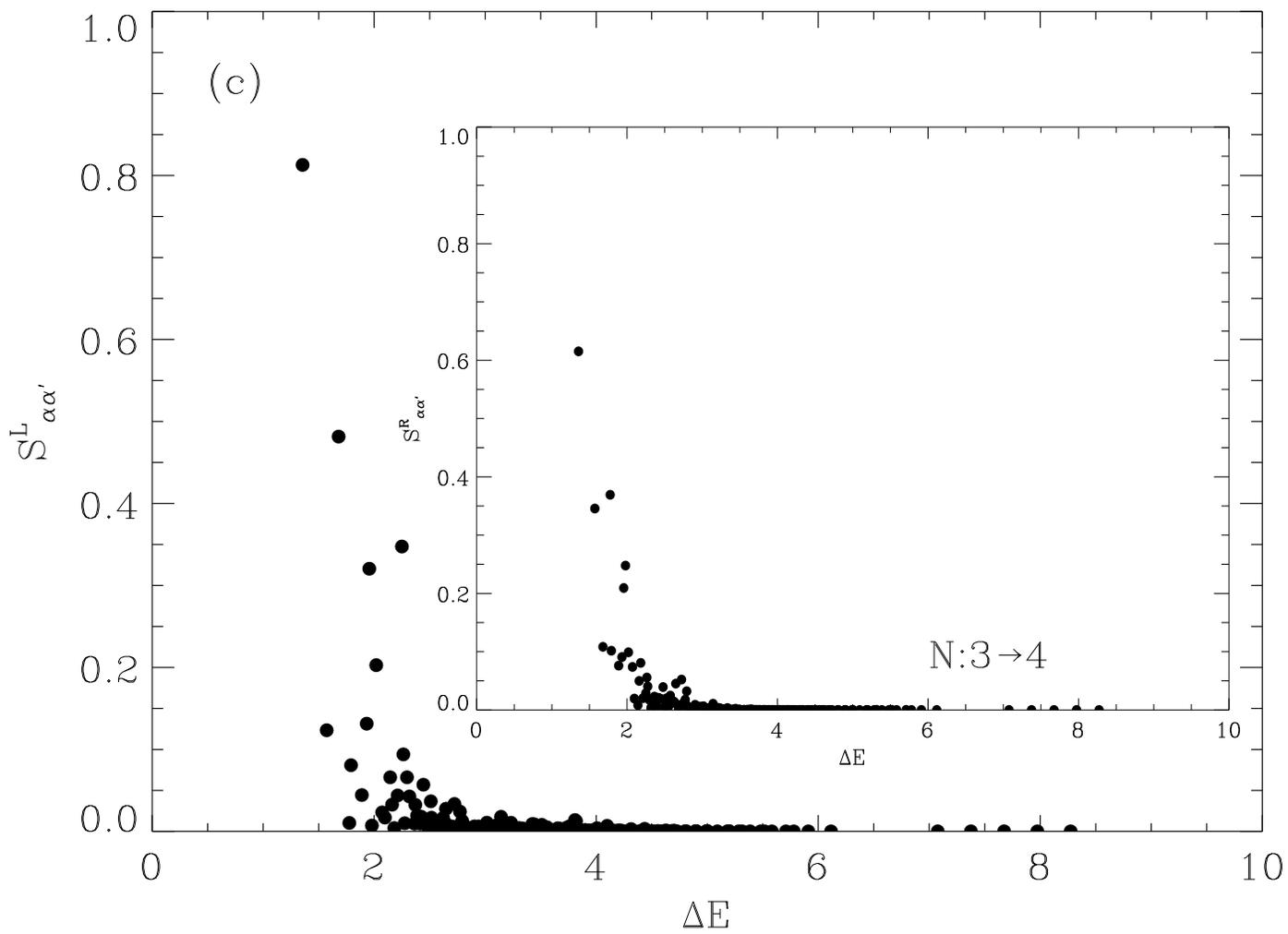
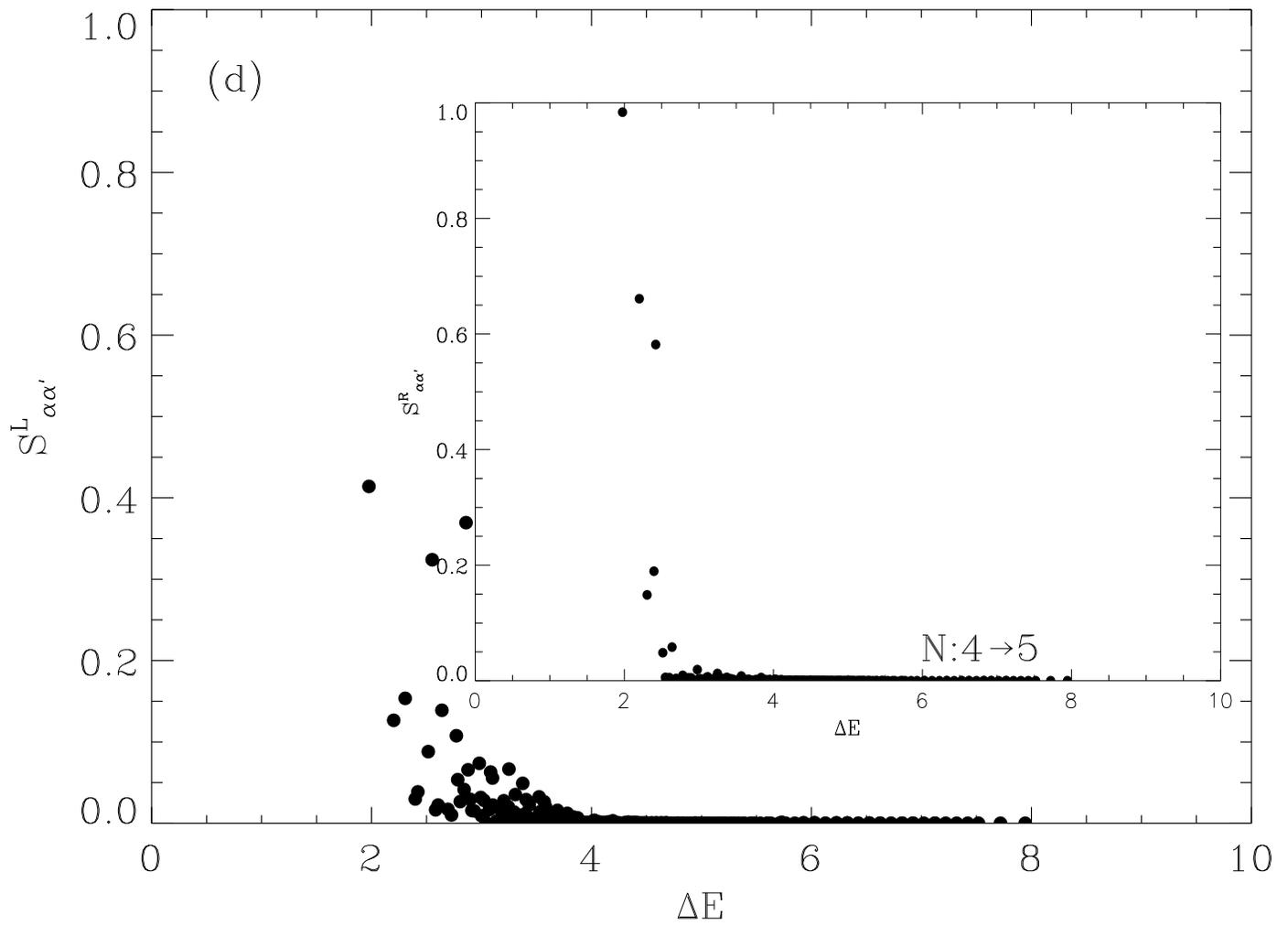
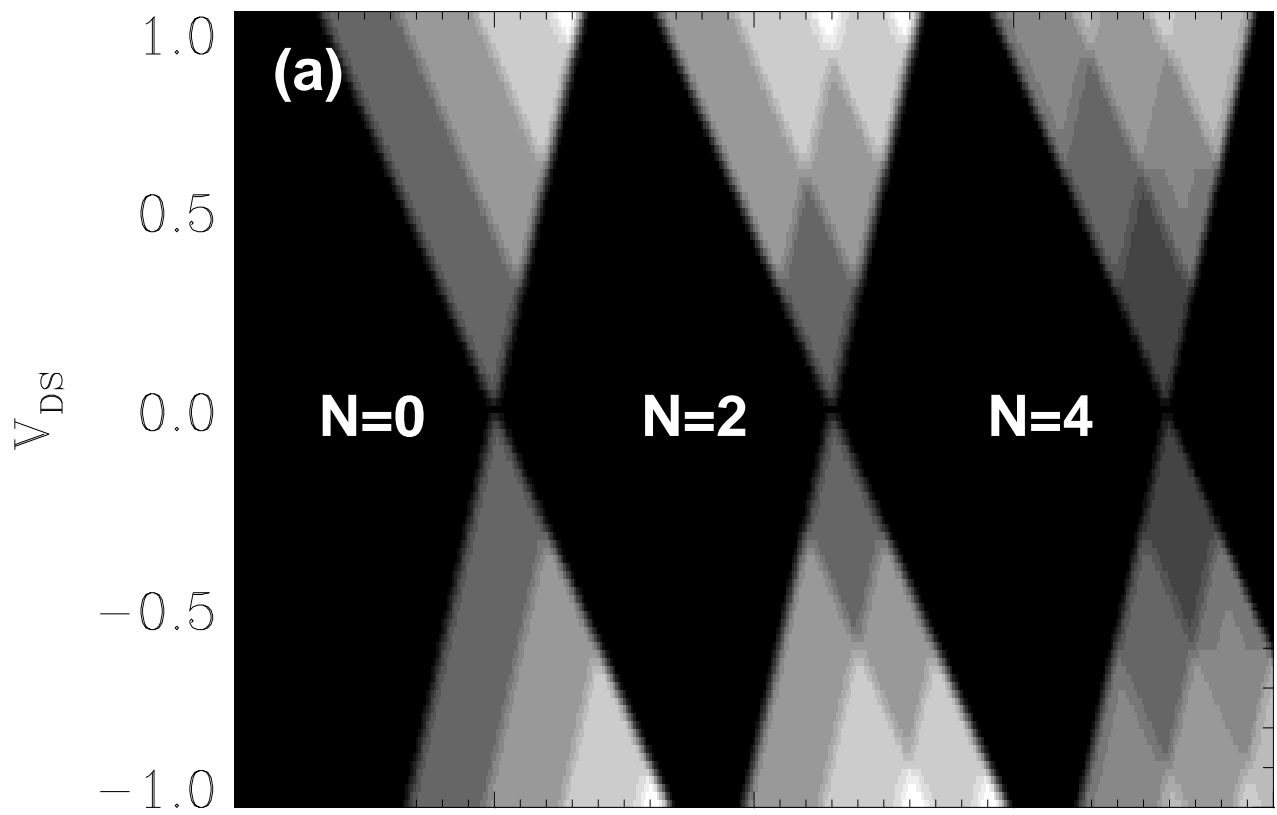
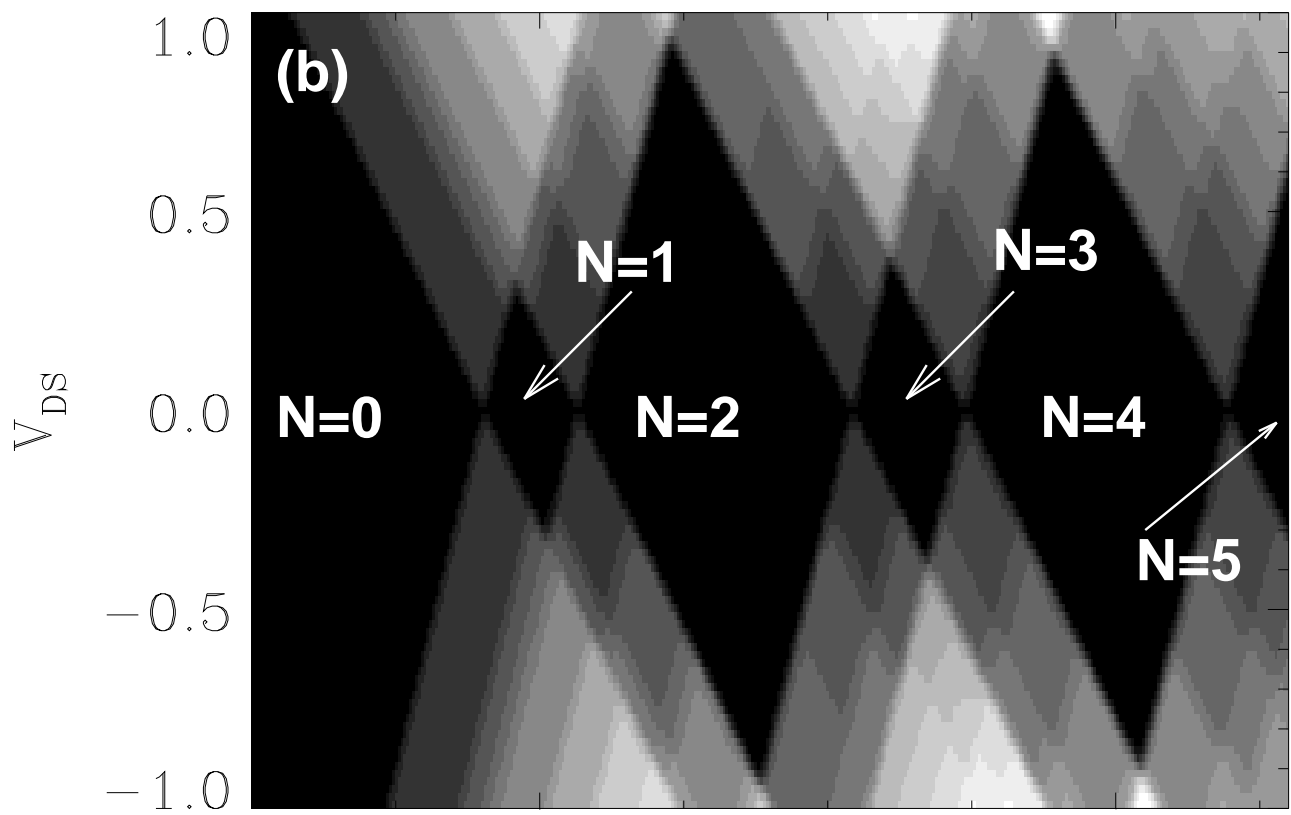
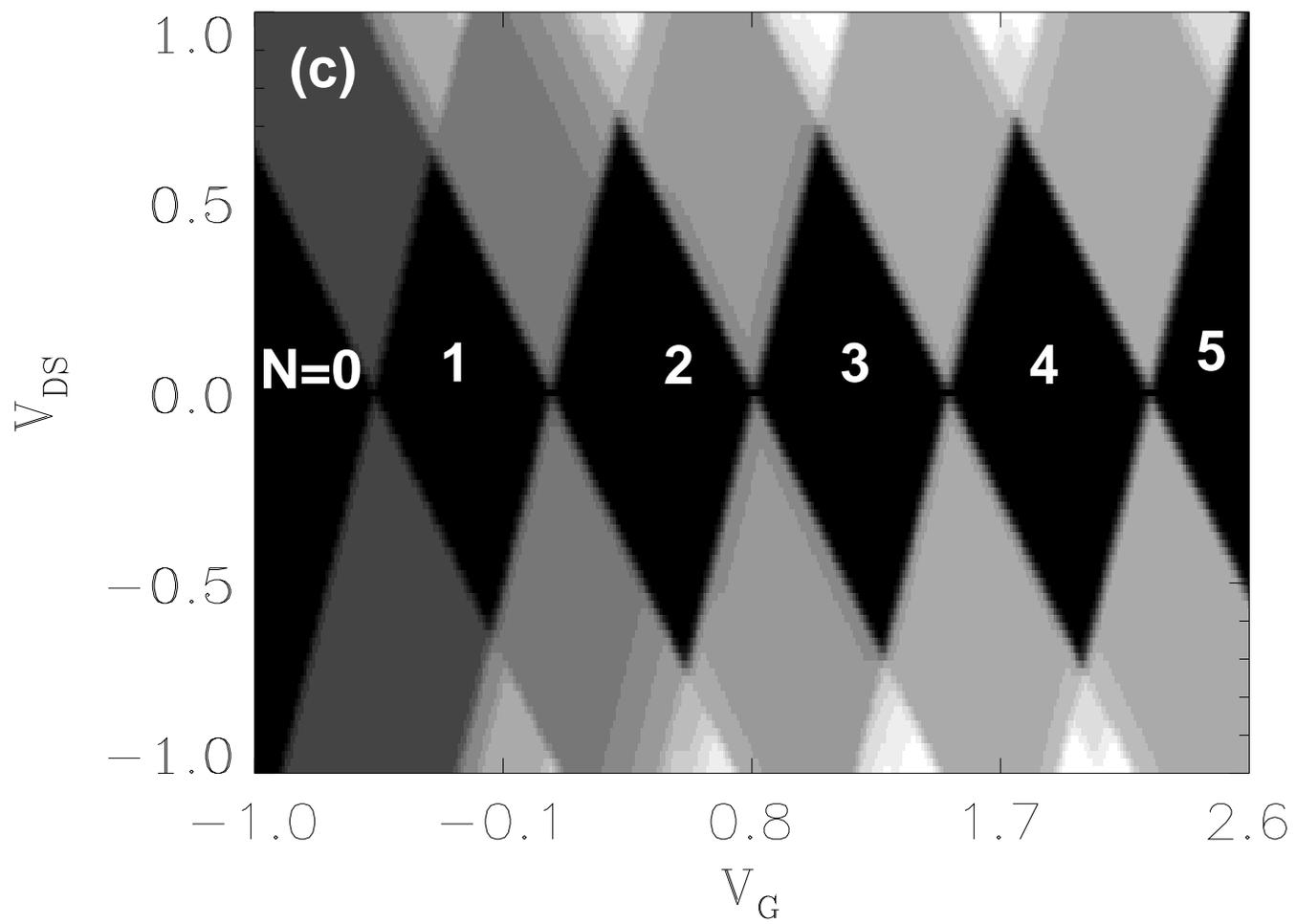
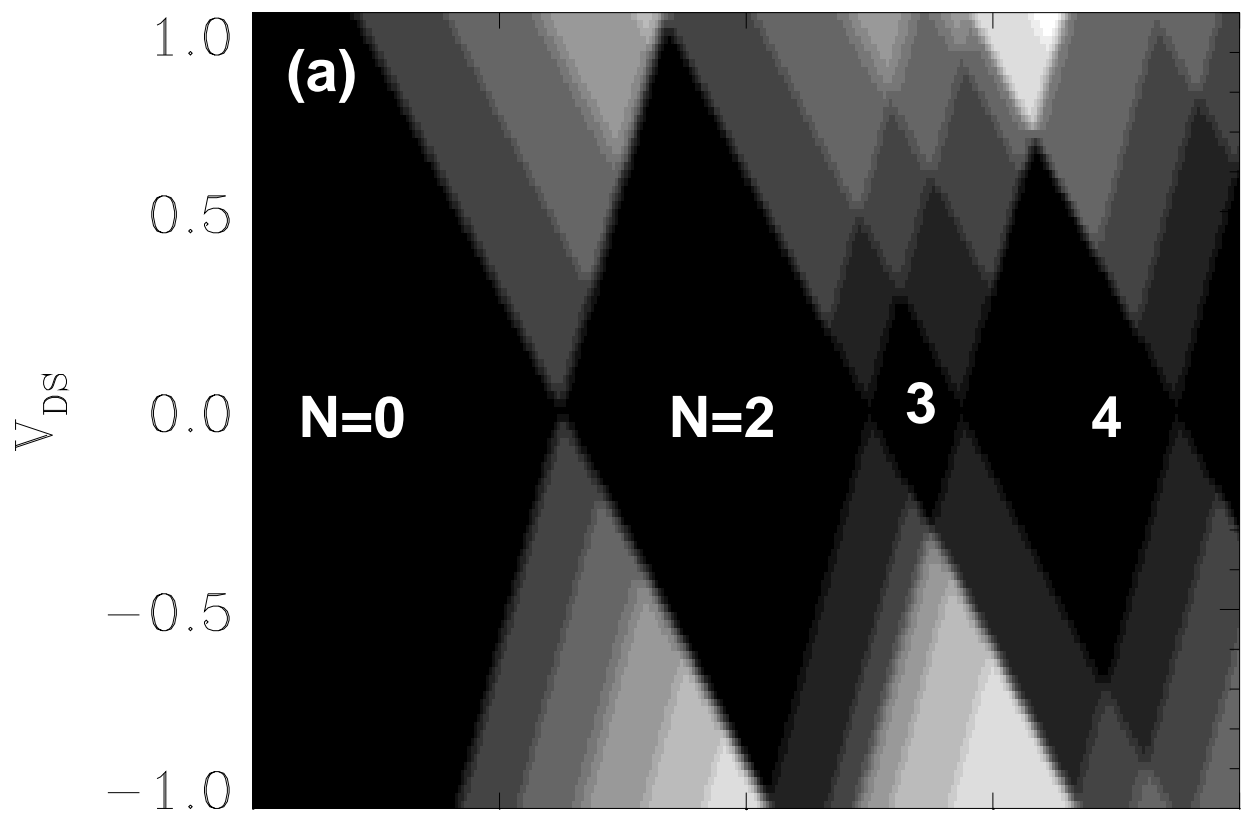
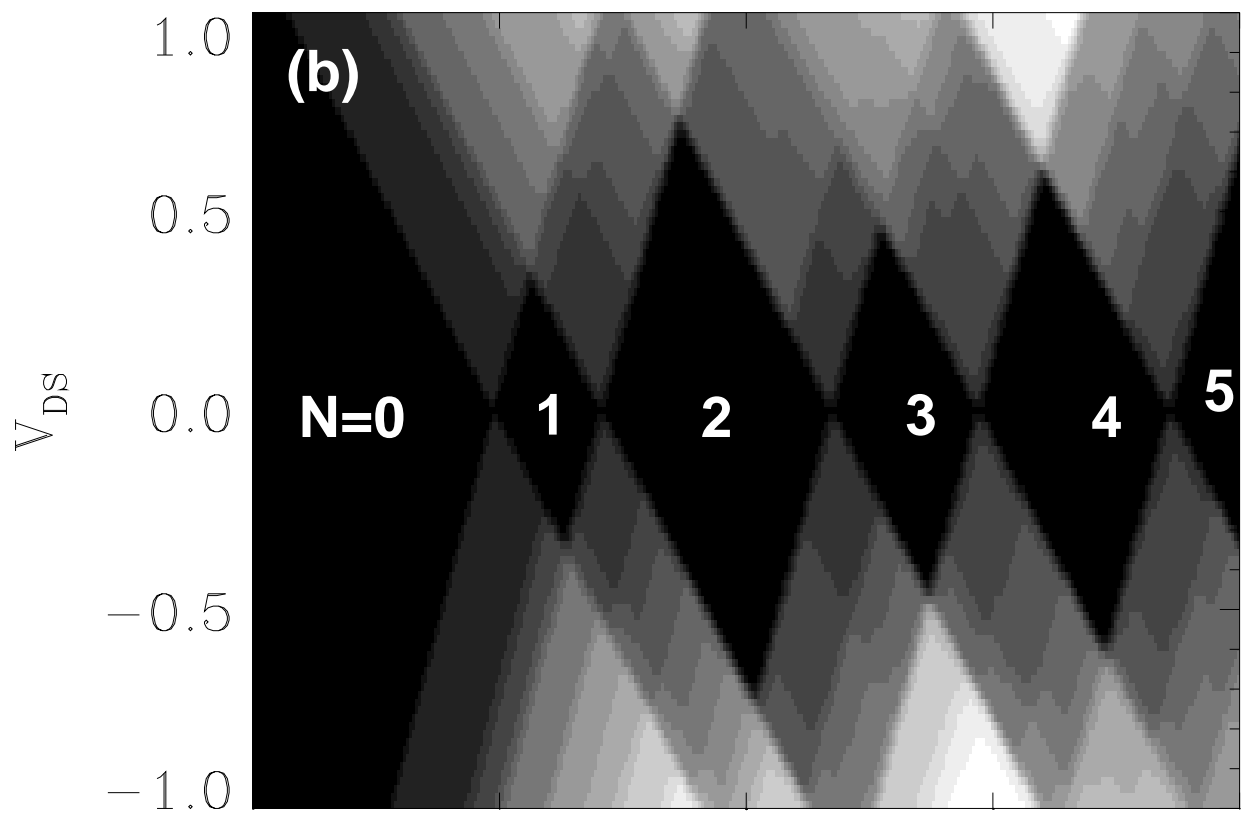
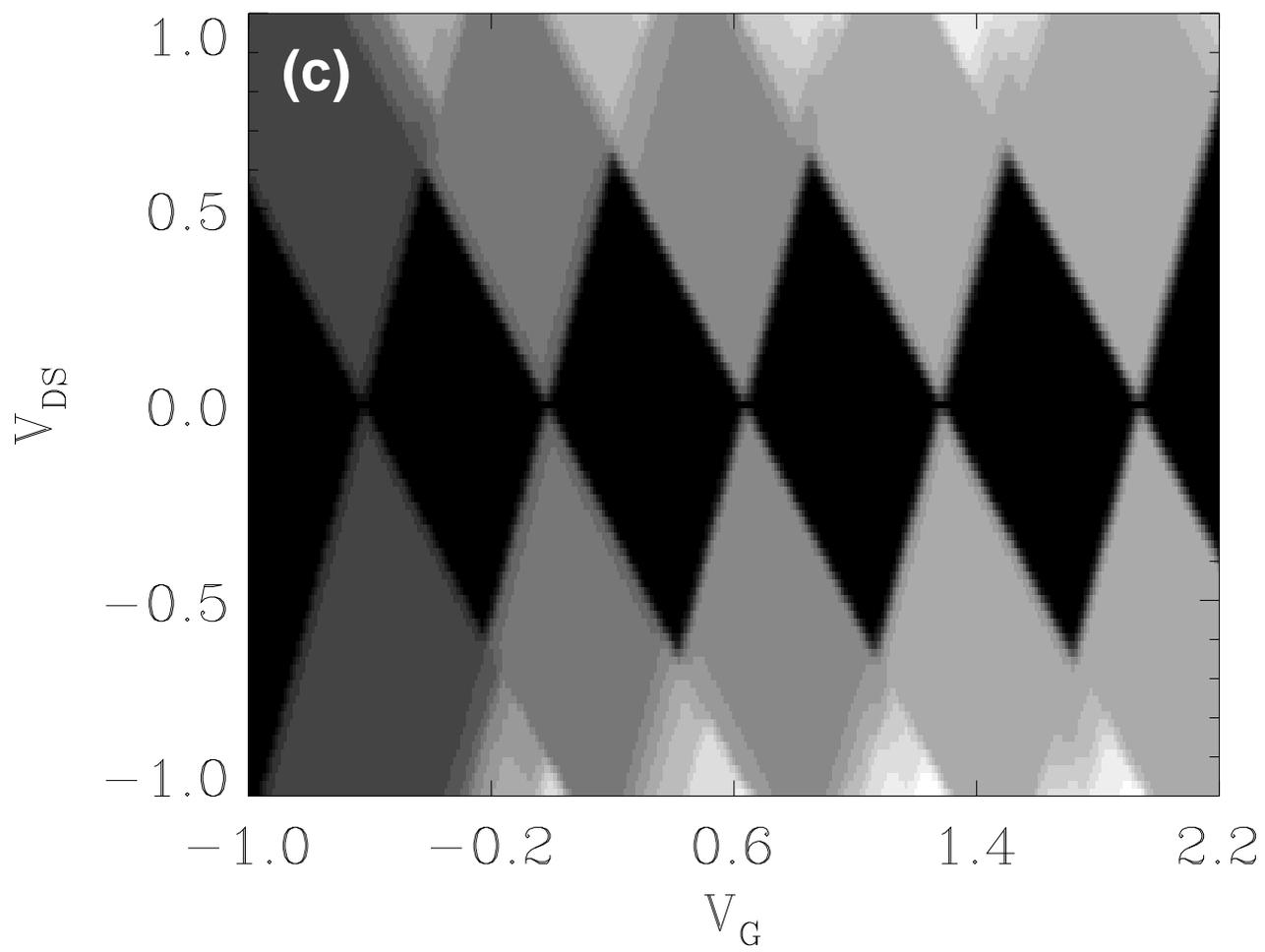
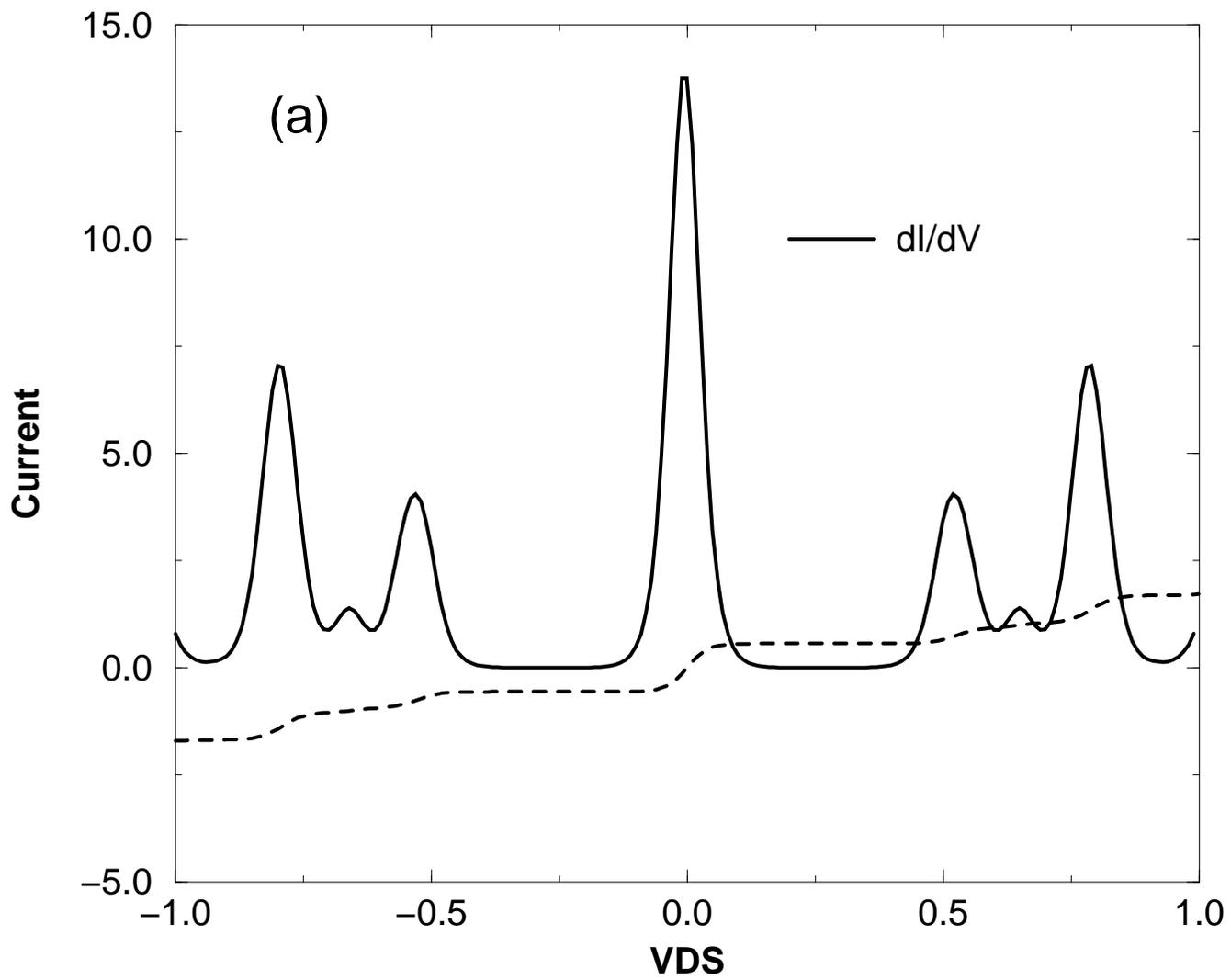
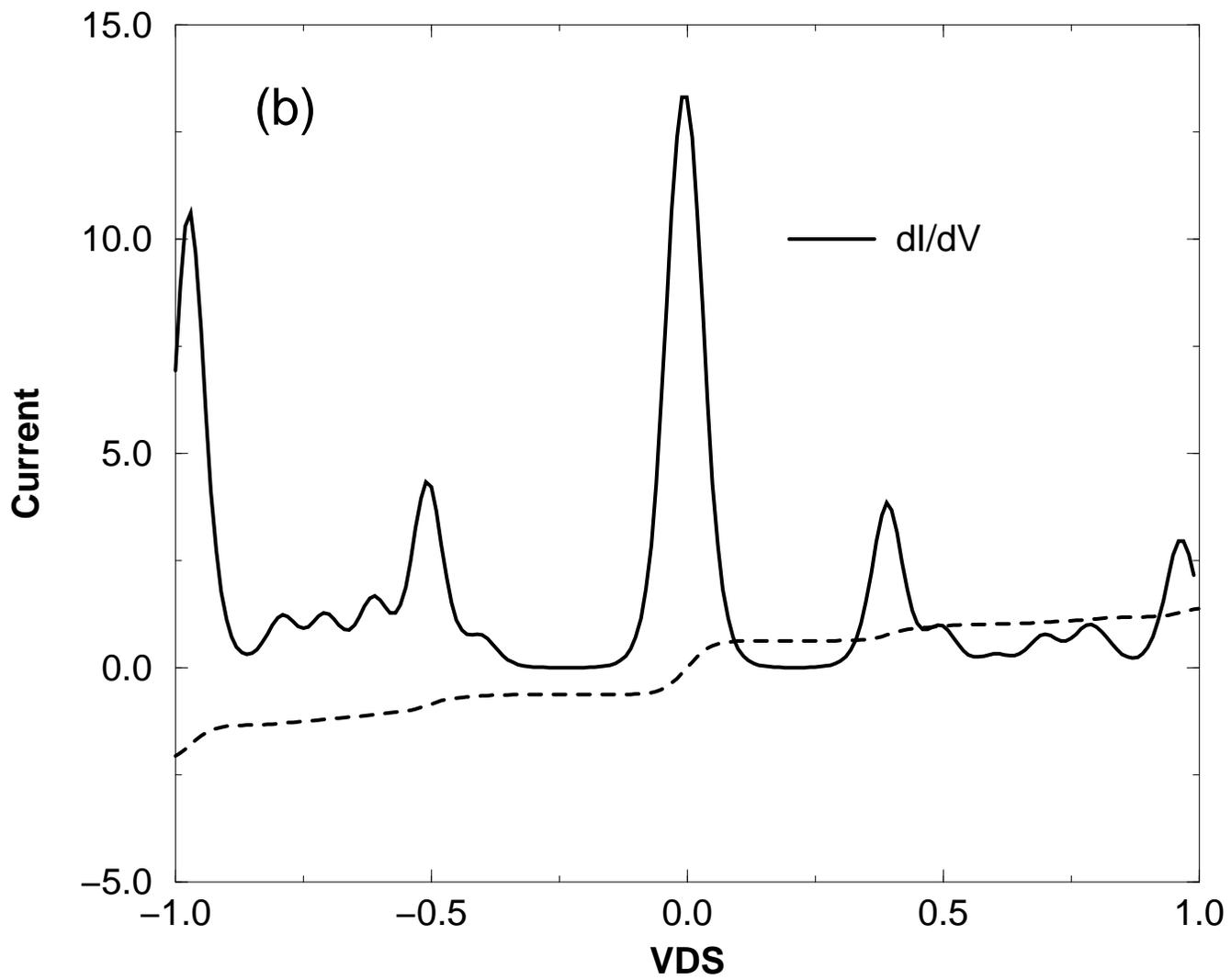
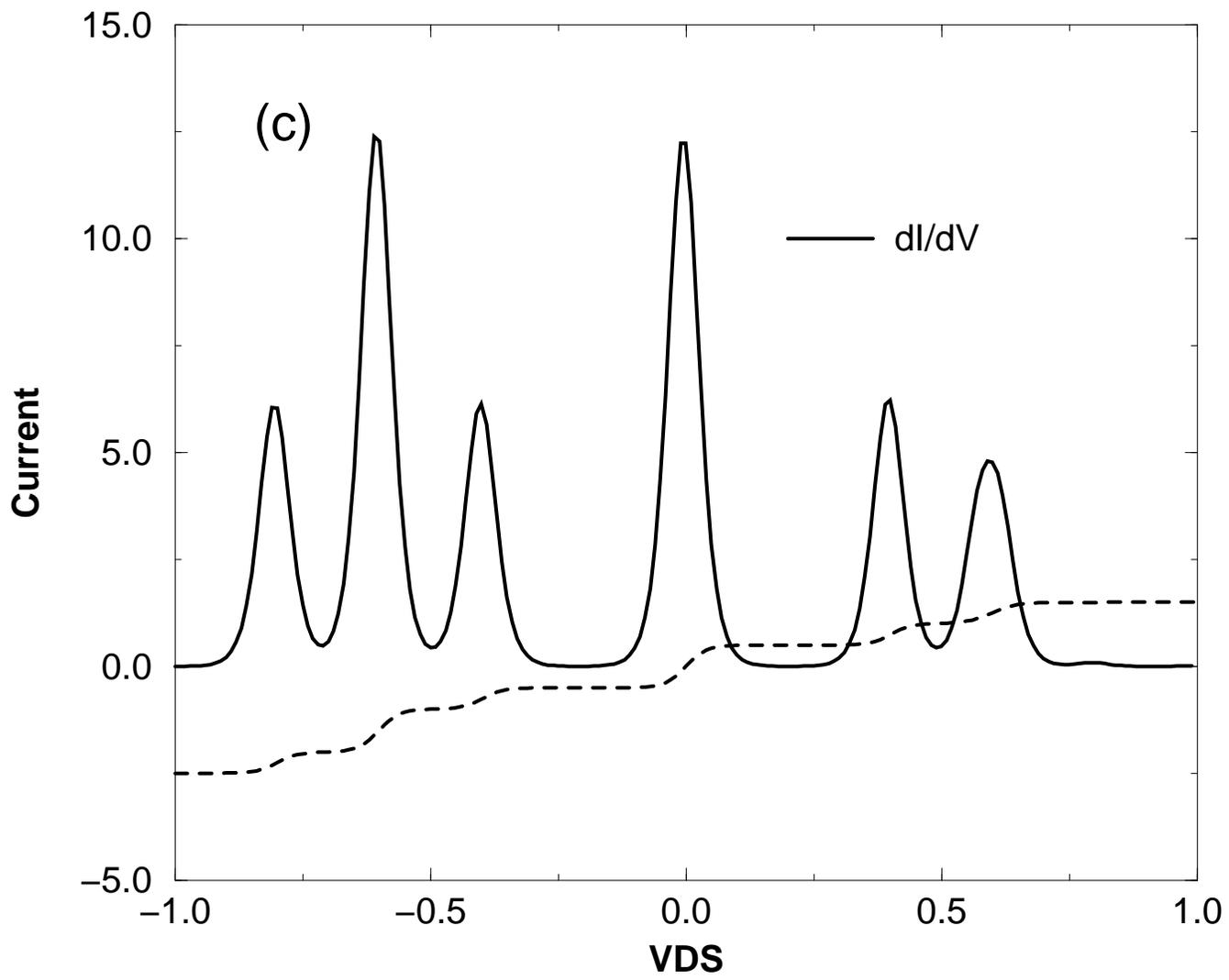
\end{document}